\documentstyle[prd,epsfig,aps]{revtex}
\def\beqra{\begin{eqnarray}}
\def\eeqra{\end{eqnarray}}
\def\beq{\begin{equation}}               
\def\eeq{\end{equation}}

\def\vp{\varphi}

\def\half{\mbox{\small$\frac{1}{2}$}}  

\def\rM{\rho_\chi}
\def\rr{\rho_{\mathrm rad}}
\def\rb{\rho_{\mathrm b}}
\def\rvp{\rho_\vp}
\begin{document}

\draft
\input epsf
\twocolumn[\hsize\textwidth\columnwidth\hsize\csname
@twocolumnfalse\endcsname

\title{Dark Energy and  Dark Matter }
\author{D. Comelli $^{(1)}$, M. Pietroni $^{(2)}$ and A. Riotto$^{(2)}$}
\address{$^{(1)}${\it INFN - Sezione di Ferrara, 
via Paradiso 12, I-35131 Ferrara, Italy  }}
\address{$^{(2)}${\it INFN, Sezione di Padova,
via Marzolo 8, I-35131, Padova, Italy}}
\date{February 2003}
\maketitle\noindent
\begin{abstract}
\noindent

It is a puzzle why the densities of dark matter and 
dark energy are nearly equal today when they scale so differently 
during the expansion of the universe. This conundrum may be 
solved if there is a  coupling between the two dark sectors. In this
paper we assume that dark matter is made of cold relics with 
masses  depending exponentially 
on the scalar field associated to dark energy. 
Since the dynamics of the system is dominated by an attractor solution, 
the dark matter particle mass  is forced to change with time
 as to ensure that  the ratio  between the energy densities of dark matter 
and dark energy
become a constant at late times and 
 one readily realizes that   the present-day dark matter abundance 
is not very sensitive to 
its value when dark matter particles decouple from the thermal  bath. 
We  show that the  dependence of the present abundance 
of cold dark matter on the parameters of the model 
differs drastically from the familiar results where no connection between
dark energy and dark matter is present.
 In particular, we analyze the 
case in which the cold dark matter particle is the lightest supersymmetric
particle.

\end{abstract}
\vskip1pc]\noindent
{\it 1. Introduction.}~~
Combined analysis of cosmological observations such as cosmic
 microwave radiation (CMB) anisotropies \cite{CMB},  cluster baryon fraction 
\cite{cluster} and supernovae Ia \cite{SN}, give increasing support for the so called `cosmic concordance' model, in 
which the universe is flat ($\Omega_{\mathrm tot} =1$), and made for  one third
 of non-relativistic dark matter (DM), and for two thirds of a smooth
component, called Dark Energy (DE). 
DM is commonly associated to weakly interacting particles (WIMPs), and can 
be described as a fluid with vanishing pressure. DE, being responsible for the accelerated expansion of the universe, 
has to be identified with some more exotic component with negative pressure, such as the cosmological constant or 
a scalar field with a proper potential \cite{DE}.
A common assumption is that there is no interaction between DM and DE. 
This implies that the ratio between their energy densities evolves in time as $\rho_{DM}/\rho_{DE} \sim a^{3W}$, 
$a$ being the scale factor of the universe and $W$ the DE equation of state ($W=-1$ for a cosmological constant, $W<-1/3$ to get acceleration). 
As a consequence, the cosmological constant or the scale of the effective potential have to be carefully tuned in order 
to get the observed proportions of DM and DE right today.

A possible solution to this `cosmic coincidence' problem may be obtained by introducing a coupling 
 between DM and DE. More specifically, if DE is identified with a scalar field $\vp$ with 
an exponential potential and the DM particle mass also depends exponentially on $\vp$, the late time behavior of the 
cosmological equations gives accelerated expansion and, at the same time, a constant DM/DE ratio \cite{Amendola}.
 This behavior relies on the existence of an {\em attractor} solution, which makes the late time cosmology insensitive 
to the initial conditions for DE and DM.
The possibility of varying mass particles (VAMPs) as DM candidates was
considered 
previously in \cite{vamps}, but with different functional dependences not 
leading to a solution of the cosmic coincidence problem.

In this letter, we will explore the impact of this scenario on the usual cold 
DM picture,
 which is based on the freeze-out at an early epoch of some non-relativistic 
species 
(see, for instance, \cite{Kolb}). In the standard scenario with constant mass 
particles the late time DM energy 
density is just the one at the freeze-out rescaled by the volume expansion.
 Instead, in the present VAMP scenario 
the attractor prompts a mass change in order to have the right DM/DE ratio 
today. As we will see, besides solving 
the cosmic coincidence problem, this will significantly  enlarge the parameter
 space allowed by the observed abundances.

\vskip1pc
\noindent
{\it 2. VAMPs and the coincidence problem}~~
We will consider as DM candidate a particle $\chi$ of mass $M$,
depending exponentially on the DE  field $\vp$,
\beq 
M_\chi(\vp)=\overline{M} \, e^{-\lambda\vp}\; ,
\eeq
where $\vp$ is expressed in units of the Planck Mass $M_p$. The scalar field
has an exponential potential
\beq
V(\vp)=\overline{V} e^{\beta \vp} \,,
\label{Vvp}
\eeq
with $\lambda, \,\beta>0$.
If the DM particle is non-relativistic, its energy density is also
$\vp$-dependent, since it is given by
\beq 
\rM=M_\chi(\vp) Y_\chi n_\gamma\;,
\label{rhoM}
\eeq
where $Y_\chi=n_\chi/n_\gamma$ is the number density of DM particles
relative to the density of photons, $n_\gamma=n_\gamma^0
(a/a_0)^{-3}$, with $n_\gamma^0=411\, {\mathrm cm}^{-3}$,
and $a/a_0$ the scale factor of the universe relative to its value
today. 

As a consequence, the scalar field is subject to an effective
potential given by the sum of Eqs.~(\ref{Vvp}) and (\ref{rhoM}).
The equation of motion is then
\beqra
&&\frac{1}{3}\frac{\rM+\rb + V+\rr}{1-\vp'^{2}/6} \vp''
+\nonumber\\
&&\half\left(\rM+\rb+2\,V+\frac{2}{3}\rr\right) \vp' =
-\beta\,V+\lambda\,\rM\;,
\eeqra
where primes denote derivatives with respect to $\xi=\log(a/a_0)$, $\rb$ and
$\rr$ are the energy densities in baryons and radiation, respectively, and we have assumed a spatially flat universe.
Since we are interested in the late-time behavior, we can assume
$\rb,\,\rr\ll \rM,\,\rvp$. In this limit it is easy to see that there
is a solution such that
\beq
\vp = \frac{-3}{\lambda+\beta}\;\xi\,,\;\;\;\;\;\;\;\;\;\;\;\;
\Omega_\vp \simeq
1-\Omega_\chi=\frac{3+\lambda(\lambda+\beta)}{(\lambda+\beta)^2}\, ,
\label{attractor}
\eeq
which is an {\em attractor} in field space if $\beta>(-\lambda+\sqrt{\lambda^2+12})/2$. From the moment
 the attractor is reached, the energy densities in DM
and in DE evolve at a constant ratio depending only on $\lambda$ and
$\beta$, thus solving  the cosmic coincidence problem. 

A closer inspection at the  solution (\ref{attractor}) using Friedman's
equations reveals that  
\beq
\frac{a\,\ddot{a}}{\dot{a}^2} = -\frac{1+3 W}{2}\,,\;\;{\mathrm
with}\;\;\;\;\;W=\frac{-\lambda}{\beta+\lambda}\,,
\eeq
that is $W$ is negative and may lead, if $W<-1/3$, to an accelerated
expansion of the universe. To
understand how it is possible to get both acceleration and constant
ratio between DM and DE one may look at  the scaling behavior of the
energy densities on the attractor (\ref{attractor}), 
\beq \rM \sim e^{-\lambda\vp-3\xi} \sim \rvp \sim e^{\beta \vp} \sim
e^{-3(W+1)\xi}\,.
\eeq
The $\vp$ dependence of the DM mass modifies the usual scaling
$e^{-3\xi}$ of non relativistic matter. Since the mass increases
with the expansion, it corresponds to an effectively negative
$W$, even though DM is still a  pressureless fluid made up by
non-interacting particles.
\begin{figure}[htb]
\epsfxsize=3.2 in \epsfbox{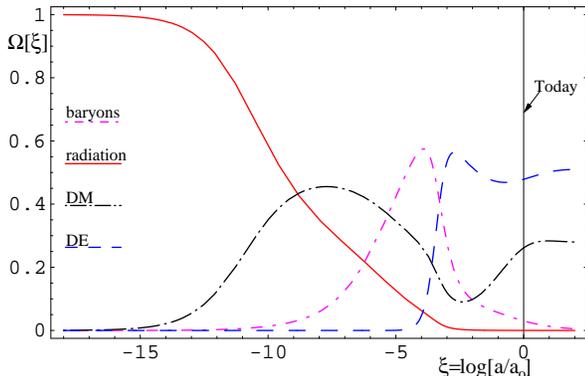}
\caption{
Time evolution of the relative abundance of different species,
 expressed as fractions of the critical density.
The corresponding parameter configuration is: 
$\beta=1.88,\,\lambda=0.82,\;\bar{V}=10^2\,{\rm GeV}^4,\; 
M_S=500\,{\rm GeV},$  $M_{\chi}^{\rm ph}=105\,{\rm GeV}$ 
 and $\alpha_S^2=0.0068$ }
\label{eom}
\end{figure}
In Fig.~\ref{eom} we plot a typical solution, including also $\rb$ and
$\rr$. During radiation domination $\rvp,\,\rM\ll\rr$, then $\vp \sim
{\mathrm const.}$, so that DM exhibits the standard scaling,
$e^{-3 \xi}$. Around matter-radiation equivalence the attractor starts
to become effective and after some transient regime -- possibly
including a baryon-dominated epoch -- it is reached at about the
present epoch. In this scenario, the future will be
characterized by values of $\Omega_\vp$, and $\Omega_\chi$ closer and
closer to the values in (\ref{attractor}), with $\rb$ and $\rr$
diluting with the usual laws, $a^{-3}$ and $a^{-4}$, respectively.

As will be discussed in \cite{AMP} the growth of fluctuations in this
scenario is compatible with the present data on CMB and large scale
structure. In particular, it is remarkable that matter fluctuations
grow even during accelerated expansion, unlike in the usual
$\Lambda$-dominated or quintessence scenario.
\vskip1pc
\noindent
{\it 2. VAMPs and DM abundance}~~
As we have discussed in the previous section, the late-time attractor
forces the DM particle mass to change with time as to ensure that DM
and DE scale at the same, constant, ratio given by
Eq.~(\ref{attractor}). In this section we show 
how this translates into an enhancement of the parameter 
space giving the observed energy densities. 
Alternative ideas  to modify  the standard freeze-out picture
of CDM suppose a kineton-dominated expansion
of the universe \cite{salati} or a low-reheating temperature
after inflation \cite{tkg}. Differently from these approaches, where
some new dynamics occurs at early epochs,
in  our proposal  the standard expansion history of the universe 
gets modified only   at late times -- after radiation-matter equality 
-- when
the field $\vp$ starts to roll.   
We assume that the DM particle
$\chi$ falls into the category of Cold Dark Matter (CDM), that
is $\chi$-particles cease annihilating during the evolution of the
universe when they were non-relativistic (
in the case of relativistic DM,
 due to the fact that the DM abundance $Y_{\chi}$ is mass independent,
 the predictions of the VAMP  and the  standard scenarios 
are the same).
This amounts to assuming that
$x_f=M_\chi\left(\varphi_f\right)/T_f >1$, where $\varphi_f$ is the
value of the DE field at the freeze-out epoch and $T_f$ is the 
freeze-out temperature.
The present DM energy density, Eq.~(\ref{rhoM}), depends on the value
of the {\it physical} $\chi$-mass today 
$M_\chi^{\rm ph} \equiv M_\chi(\vp_0)\equiv \overline{M} e^ {-\lambda
\vp_0}$ (where $\vp_0$ is the value of the DE field
today) and on the comoving number density $Y_\chi$. The latter
depends,
among other parameters, on the $\chi$ mass at the earlier epoch of
freeze-out, {\it i.e.} on $M_\chi(\vp_f)\equiv \overline{M} 
e^{-\lambda \vp_{f}}=M_\chi^{\rm ph}e^{\lambda \Delta \vp}$, 
where $\Delta \vp \equiv \vp_0-\vp_{f}$. 

In order to be definite, we will consider the case of stable 
$\chi$-particles annihilating into light states 
by exchanging an heavier particle $S$ of mass $M_S$, 
 $\chi\overline{\chi}\rightarrow$ light states. To estimate $x_f$ we make 
use of the
criterion $\Gamma_{\rm ann}(x_f)\simeq H(x_f)$, where $\Gamma_{\rm ann}=
n^{\rm eq}_\chi\langle\sigma_{\rm ann}|v|\rangle$ and 
$\langle\sigma_{\rm ann}|v|\rangle$ is the thermally-averaged annihilation
cross-section times velocity of the DM particles. 
We also suppose that
the DM particles annihilate through a  $p$-wave suppression channel.
 This happens --
for instance --  if $\chi$
is a Majorana fermion. 

\begin{figure}[htb]
\epsfxsize=3.2 in
\epsfysize=2.8 in
 \epsfbox{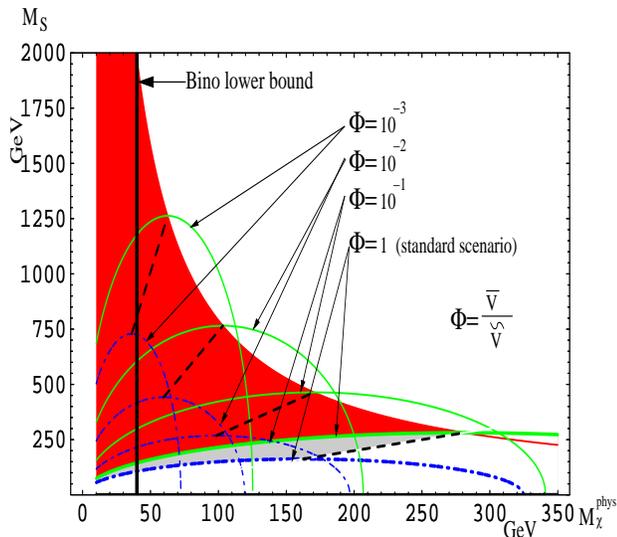}
\caption{
Regions with  $0.1 \leq 
\Omega_{\chi}h^2 \leq 0.3$
and  $M_S > M_{\chi}$  for fixed values of 
$\Delta \vp$, see Eq.~(\ref{scaling}).  
The grey region represents the parameter space available in
the standard scenario  $\Delta \vp=0$.
The red+grey region represents the  total parameter space  
available in the VAMP scenario.
}
\label{contours}
\end{figure}

Under these assumptions, the thermally-averaged 
annihilation cross section at freeze-out  is given by
\beq
\langle \sigma_{\mathrm ann}|v|\rangle
 = \alpha_S^2\,\frac{T_f}{M_\chi(\vp_f)}
\frac{M_\chi^2(\vp_f)}{(M_\chi^2(\vp_f)+M_S^2)^2}\equiv \sigma_0\,
\frac{T_f}{M_\chi(\vp_f)}\, ,
\label{sigma}
\eeq
where $\alpha_S$ is a combination of coupling constants
measuring the strengths of the interactions.
The standard freeze-out picture \cite{Kolb} gives the number
abundance at late times $
Y_\chi(T_0)\simeq 
Y_\chi(T_f) = (7.6 x_f^2/g_*^{1/2}(T_f)\,M_{p}\,
 \sigma_0\, M_\chi(\vp_f))$, 
where $T_0\simeq 2.75$ K is the present-day photon temperature
and $g_*$ counts the relativistic degrees of freedom in the plasma
at the freeze-out epoch.
The energy density of the DM particles today reads 
\beq \label{omega}
\Omega_\chi(t_0) h^2=c\,e^{-3\lambda \Delta\vp}
\frac{\, \left(e^{2 \lambda \Delta\vp} \left(M_\chi^{\rm ph}\right)^2
+M_S^2\right)^2}{ \left(M_\chi^{\rm ph}\right)^2}
\eeq
with $ c\simeq 1.46\times 10^{-10}\alpha_S^{-2}
 (x_f^2/g_*^{1/2}(T_f))\,{\rm GeV}^{-2}$.
The standard $\Lambda$CDM scenario is recovered by setting $\Delta \vp=
0$ or, equivalently, $\lambda=\beta=0$. In this case,
$\overline{V}$  plays the role of the cosmological constant, $M_\chi=
M_\chi^{\rm ph}$ at any epoch, and constant
$\Omega_\chi(t_0) h^2$  
contours in the $(M_\chi-M_S)$ plane are those corresponding to $\Phi=1$ in
Fig.~\ref{contours}, where the region $M_S<M_\chi$  
has been excluded.
 It is worth while stressing that, once the
acceptable range for $\Omega_\chi(t_0) h^2$ has been chosen, the DE
abundance has to be fine tuned to give a  flat universe (at, say, ten
percent accuracy). This translates into a constraint on $\overline{V}$
which is nothing but the cosmic coincidence problem of the $\Lambda$CDM scenario.

In the VAMP scenario the situation changes dramatically. Keeping the
physical abundance,
$\Omega_\chi(t_0) h^2$, fixed and assuming the attractor has been reached
by today, the circular contour for $\Delta \vp=0$ is deformed to a ellipse of
the same area and eccentricity given by $e^{\lambda \Delta \vp}$, as
in Fig.~\ref{contours}. On the same contour the potential energy of
 the $\vp$ field
is also constant, $\overline{V} e^{\beta \vp_0}$, so  we can express
$\Delta \vp$ in terms of $\overline{V}$ as
\beq \label{dvp}
\Delta \vp=-\frac{1}{\beta}\log\frac{\overline{V}}{\widetilde{V}} 
\equiv -\frac{1}{\beta}\log\Phi\;,
\label{scaling}
\eeq
where $\widetilde{V}$ is the particular value of $\overline{V}$ such that
$\Delta \vp=0$. 

Compared to the $\Lambda$CDM case, in the VAMP case we can vary
$\overline{V}$ by many orders of magnitude,  see Fig.~\ref{contours}. In principle, any value of $\overline{V}$ would give an acceptable ellipse.
In practice, extreme values would prevent the field
$\vp$ from reaching the attractor before the present epoch, thus deforming the corresponding contour from an elliptical shape and reducing its area. 
However,
as we have checked numerically, the allowed values for $\overline{V}$ 
typically span  many orders of magnitude, thus removing any fine
tuning on this parameter.

The most striking feature is that observationally
allowed values of $\Omega_\chi(t_0) h^2$ can be obtained
for very large masses of the scalar particle $S$. 
In the traditional
$\Lambda$CDM scenario large scalar masses
would lead to a suppression of the annihilation cross-section and therefore
to a over-abundance of DM particles today. On the
other hand, in the case in which
the $\chi$-mass depends on the DE field, the initially large 
value of the $\chi$-energy density  at the freeze-out is adiabatically 
reduced not only by the expansion of the universe, but also by the
fact that, due to the running
of the DE field, the $\chi$-mass is smaller today
 than  at freeze-out. For consistency, we have to impose in this case that $\chi$ has always been lighter than $S$, which is ensured by the constraint $M_\chi(\vp_f)<M_S$, giving the straight dashed lines in Fig.~2.


Even more remarkable from a phenomenological point of view is the drastic
enhancement of the allowed parameter space in the $(M_\chi^{\rm ph}-M_S)$
plane. Varying $\overline{V}$, the allowed regions span an area
delimited by an hyperbole, which contains values for $M_S$ much larger than in the standard scenario. 

Points above the hyperbole give still the good DM/DE ratio, but fail in reproducing their absolute value, that is, the Hubble parameter today.

\vskip1pc
\noindent
{\it 4. An illustrative example: supersymmetric dark matter.}~~
Of the many CDM candidates, the best motivated from a particle
physics point of view  is the  supersymmetric neutralino 
if the latter is the lightest supersymmetric particle (LSP) \cite{dmreview}.
Most of the supersymmetric models obtained from supergravity usually predict
that the LSP is an almost pure $B$-ino. Therefore, 
to give an illustrative example of our previous findings, we consider 
the case in which the  LSP is a 
$B$-ino and its mass $M_B$ depends exponentially upon the time-dependent 
DE field
$\phi$, $M_B=\overline{M}_B\,{\rm exp}\left(-\lambda\vp\right)$.

Being the $B$-ino the supersymmetric partner of the
 $U(1)_Y$ gauge boson, it does not directly contribute at the
one-loop level to the running  of  gauge couplings. This means that the
time-dependence of the electromagnetic coupling is induced by the bino mass only
through tiny  quantities parametrizing 
how much the LSP composition
depends on the Wino and Higgsinos. A full parameter space analysis regarding this
specific point will be presented in \cite{noi}.
In the early universe, $B$-inos  mainly annihilate into fermion
pairs through $t$-channel exchange of squarks and sleptons.
Exceptions occur only for pathological situations in which there is a
resonant $s$-channel exchange of $Z^0$ or a Higgs boson.
Because of the large hypercharge of the right-handed electron and the
expected lightness of sleptons compared to squarks, it is often a good
approximation to include in the annihilation cross section only the
exchange of the right-handed sleptons. 

Summing over three slepton
degenerate families with mass ${\tilde M}_{\ell_R}$, the thermally-averaged 
$B$-ino
annihilation cross section times velocity has the same form as eq.~(\ref{sigma}), with $M_S$ and $M_\chi$ replaced by ${\tilde M}_{\ell_R}$ and $M_B$, respectively, and 
$\alpha^2_S=  24\pi\alpha^2/\cos^4\theta_W\,\simeq 6.7\times10^{-3} $.

In the traditional case in which the mass of the $B$-ino is not 
dependent upon the DE field, $M_B=M_B^{\rm ph}$,  
cosmological considerations give an upper bound to the $B$-ino
mass. Indeed, the requirement that charged particles are not the LSP
implies ${\tilde M}_{\ell_R}>M_B$. The minimum allowed $B$-ino relic
abundance corresponds to the maximum annihilation cross section and
therefore to the minimum ${\tilde M}_{\ell_R}$. Setting ${\tilde
M}_{\ell_R} =M_B$ in the expression for $\Omega$, one obtains an upper
bound on the $B$-ino mass of about 300 GeV (for $\Omega h^2<0.3$)
\cite{bees}.  This bound can be weakened in several ways. For example 
through the presence of 
resonant $s$-channel annihilations, once a small Higgsino admixture
is introduced. Furthermore, as emphasized in Ref.\ \cite{falk}, 
whenever the sleptons and
the $B$-ino become degenerate in mass within about 10--20{\%}, one
cannot ignore the effects of coannihilation. These effects can modify
significantly the $B$-ino relic abundance, because annihilation
channels involving the charged sleptons have large cross sections
which are not $p$-wave suppressed. Indeed, even in the case of the
constrained model, the previous limit on the $B$-ino mass can be
relaxed to about 600 GeV \cite{falk}.
Coannihilation effects do not significantly modify the bound on the
slepton mass for a fixed value of $M_B$ (as long as it is not too
close to ${\tilde M}_{\ell_R}$).

On the other hand, these bounds on the slepton and $B$-ino masses
can be drastically modified if the mass of the 
$B$-ino depends upon the DE field. These effects are 
illustrated in Fig.~ \ref{contours} at different values of the parameter
$\overline{V}$. As expected from the findings of the previous section, 
observationally
allowed values of $\Omega_B(t_0)\,h^2$ can be obtained  
for very large slepton masses for given values of $\overline{V}$: 
the smaller $\overline{V}$ is,  the larger the slepton mass can be. 
Furthermore, allowing $\overline{V}$ to vary enhances drastically the allowed 
parameter space in the $(M_B^{\rm ph}-{\tilde
M}_{\ell_R})$
plane.
We leave a  more complete  analysis of the parameter
space of the supersymmetric models for a future publication \cite{noi}.
It is intriguing that a coupling between the dark energy and the dark
matter sectors may provide 
not only a natural  solution to the coincidence problem 
but also a drastic modification of the predicted abundance of CDM 
in terms of the particle physics parameters.

{\it Acknowledgements.}~~We thank Antonio Masiero 
for useful discussions. This work was
partially supported by European Contracts HPRN-CT-2000-00148 and
HPRN-CT-2000-00149.


\begin{references}
\bibitem{CMB} P.~de Bernardis {\it et al.},
Astrophys.\ J.\  {\bf 564}, 559 (2002); R.~Stompor {\it et al.},
Astrophys.\ J.\  {\bf 561}, L7 (2001).

\bibitem{cluster} J.~E.~Carlstrom {\it et al.},
astro-ph/0103480.

\bibitem{SN}A.~G.~Riess {\it et al.}  [Supernova Search Team Collaboration],
Astron.\ J.\  {\bf 116}, 1009 (1998);
S.~Perlmutter {\it et al.}  [Supernova Cosmology Project Collaboration],
Astrophys.\ J.\  {\bf 517}, 565 (1999).

\bibitem{DE} For reviews on DE see for instance S.~M.~Carroll,
Living Rev.\ Rel.\  {\bf 4}, 1 (2001); P.~J.~Peebles and B.~Ratra,
astro-ph/0207347.

\bibitem{Amendola} L.~Amendola, Phys.\ Rev.\ D {\bf 62} (2000) 043511; L.~Amendola and D.~Tocchini-Valentini,
Phys.\ Rev.\ D {\bf 66}, 043528 (2002); M.~Gasperini, F.~Piazza and G.~Veneziano,
Phys.\ Rev.\ D {\bf 65} (2002) 023508;
M.~Pietroni,
hep-ph/0203085.
\bibitem{vamps} J.~A.~Casas, J.~Garcia-Bellido and M.~Quiros,
Class.\ Quant.\ Grav.\  {\bf 9}, 1371 (1992);
G.~W.~Anderson and S.~M.~Carroll,
astro-ph/9711288.

\bibitem{Kolb} E.W.~Kolb and M.S.~Turner, "The Early Universe",
 Addison-Wesley, 1990.

\bibitem{AMP} L.~Amendola, S.~Matarrese, and M.~Pietroni, in preparation.

\bibitem{salati}  P. Salati, astro-ph/0207396; 
M. Joyce, T. Prokopec, JHEP  0010:030 (2000); 
M. Joyce, Phys. Rev. {\bf D55} 1875 (1997).

\bibitem{tkg}
G.~F.~Giudice, E.~W.~Kolb and A.~Riotto,
Phys.\ Rev.\ D {\bf 64}, 023508 (2001);
N.~Fornengo, A.~Riotto and S.~Scopel,
hep-ph/0208072.





\bibitem{dmreview} K.~A.~Olive,
astro-ph/0301505.

\bibitem{noi} D. Comelli, N. Fornengo, M. Pietroni and A. Riotto, in
preparation.


\bibitem{bees} K. A. Olive and M. Srednicki, Phys.  Lett. 
{\bf B230}, 78 (1989).

\bibitem{falk} J. Ellis, T. Falk, and K. A. Olive, Phys.  Lett.  
B444, 367 (1998); J. Ellis, T. Falk, K. A. Olive, and 
M. Srednicki, Astropart. Phys. 13, 181 (2000). For recent studies, see
J.~R.~Ellis, T.~Falk, K.~A.~Olive and M.~Srednicki,
Astropart.\ Phys.\  {\bf 13}, 181 (2000)
[Erratum-ibid.\  {\bf 15}, 413 (2001)]; 
T.~Nihei, L.~Roszkowski and R.~Ruiz de Austri,
JHEP {\bf 0207}, 024 (2002).






\end{references}
\end{document}